# Electronic properties of nickelate superconductor $R_3Ni_2O_7$ with oxygen vacancies

Xuelei Sui[1,2], Xiangru Han[2], Xiaojun Chen[3], Liang Qiao[4,3], Xiaohong Shao[1*], and Bing Huang[2,5*]

[1] College of Mathematics and Physics, Beijing University of Chemical Technology, Beijing 100029, China

[2] Beijing Computational Science Research Center, Beijing 100193, China

[3] School of Mechanical Engineering, Chengdu University, Chengdu 610106, China

[4] School of Physics, University of Electronic Science and Technology of China, Chengdu 611731, China

[4] Department of Physics, Beijing Normal University, Beijing 100875, China

E-mails：shaoxh@buct.edu.cn, Bing. Huang@csrc.ac.cn

## Abstract

The discovery of superconductivity in $La_3Ni_2O_7$ has attracted significant research interest in the field of nickelate superconductors. Despite extensive studies on pristine $La_3Ni_2O_7$, the impact of oxygen vacancies ($V_O$), a common type of intrinsic defect in oxides, on electronic structures and superconductivity in $La_3Ni_2O_7$ remains unclear. In this article, we identify the most energetically favorable location for $V_O$ formation as the oxygen atom connecting the $NiO_6$ bilayer, resulting in a significant reduction in the lattice constant along the $c$-axis. Interestingly, the electronic structure undergoes notable changes, particularly for the Ni $d_{z^2}$ and Ni $d_{x^2-y^2}$ orbitals. The Ni $d_{z^2}$ orbitals change from partially filled in the pristine $La_3Ni_2O_7$ to completely filled in the presence of $V_O$, leading to a considerable decrease of its proportion near the Fermi level. Conversely, the proportion of Ni $d_{x^2-y^2}$ states increases due to the orbital localization and slight upward shift. Additionally, we observe a significant increase in the hopping of intra-bilayer Ni $d_{z^2}$ orbitals when the $V_O$ exists, but with an opposite sign, which differs greatly from the previous understanding. The inter-orbital hopping between Ni $d_{z^2}$ and Ni $d_{x^2-y^2}$ orbitals also changes its sign in the presence of $V_O$. Our results indicate that the formation of $V_O$ may be harmful to the superconductivity in $La_3Ni_2O_7$, given the general assumption for the critical role of Ni $d_{z^2}$ in generating superconductivity. Furthermore, we suggest that $Ce_3Ni_2O_7$, which shares similar electronic structures to $La_3Ni_2O_7$ but has a larger lattice volume, may be a better candidate for nickelate superconductor due to its lower $V_O$ concentration.

**Introduction.**

Besides of the infinite-layer nickelates $R$NiO$_2$ ($R$ = La, Nd, or Pr) [1-5], Ruddlesden-Popper bilayered perovskite La$_3$Ni$_2$O$_7$ (La327) under high pressure is recently discovered to be another nickelate-based unconventional superconductor [6-10]. Unlike the low superconducting temperatures observed in $R$NiO$_2$ thin films (e.g., ~10 K at ambient pressure [1] and ~30 K at 12.1 GPa [11]), bulk La327 exhibits significantly higher $T_c$ values, reaching up to ~80 K at 14 GPa [6-8]. Importantly, both monocrystalline and polycrystalline samples are found to be superconducting which undergo a structural phase transition from Amam to Fmmm around 14-15 GPa [8,12]. For the basic electronic structure, it is found that with the NiO$_6$ bilayers La327 possesses a Ni $3d^{7.5}$ configuration with fully occupied $t_{2g}$ orbitals, bonding-antibonding molecular $d_{z2}$ states, and nearly quarter-filled itinerant $d_{x2-y2}$ states [6,13]. The developed effective models suggest that the significant intra-orbital exchange of Ni-$d_{z2}$ orbitals, as well as the inter-orbital hybridization between Ni $d_{z2}$ and $d_{x2-y2}$ orbitals through O-$p$ orbitals, may play a crucial role in the observed superconductivity [14-22]. Meanwhile, the paring symmetry [23-25], the role of the Hund coupling [14,25,26] and the electronic correlation effects of pristine La327 [27-30] are extensively studied. Until now, the significantly different features compared to infinite-layer nickelates are observed in La327, although the superconductivity mechanism in La327 have not reached an agreement.

In La327 samples, the existence of oxygen vacancies (V$_O$) is inevitable [6,7,9], similar to the situations in many other oxides [35-41]. The experimental investigations have revealed that a significant concentration of V$_O$ (La$_3$Ni$_2$O$_{7-\delta}$, where $\delta$ > 0.08) may induce phase transitions and even trigger metal-semiconductor transitions under ambient pressure conditions [42-44]. Additionally, certain samples with a higher V$_O$ concentration, e.g., $\delta$ = 0.65, can exhibit weak ferromagnetism [42]. Particularly, the recently discovered uniformed resistivity in La327 under pressure may be highly related to the presence of oxygen defects [6-10]. To gain a deeper understanding of the relationship between V$_O$ and the electronic properties of La327, it is crucial to investigate the energetically favorable locations of these vacancies and their influences on the electric structures of La327 under varying pressure conditions, which is, however, still lacking. Furthermore, it is valuable to consider the presence of V$_O$ in various rare-earth bilayered perovskites [45], given the fact that different rare-earth elements in infinite-layer nickelates and cuprates can also be realized [34,46,47]. Therefore, the major purpose of this study is to systemically identify the role of Vo in La327 and suggest a possible way via rare-earth replacement to enhance the superconductivity in $R$327.

In this article, using first-principles calculations, we systematically investigate the effect of V$_O$ in the structures and electronic structures of La327. Our calculations reveal that the apical octahedral oxygen sites connecting NiO$_6$ bilayers are the most energetically favorable sites for generating V$_O$. The introduction of apical V$_O$ induces significant changes in the Ni $d_{z2}$ orbital. Specifically, the Ni $d_{z2}$ orbitals of the Ni atoms connected to the V$_O$ become fully occupied, resulting in a substantial proportion decrease at the Fermi level. Furthermore, the hopping parameters for the $d_{z2}$-$d_{x2-y2}$ two-orbital model are calculated. The intra-orbital coupling between intra-layer Ni $d_{z2}$ in defective La327 is found to be much larger but with opposite sign compared to that in pristine La327. The inter-orbital hopping between Ni $d_{z2}$ and Ni $d_{x2-y2}$ undergoes a sign change when the V$_O$ exists. In addition, our calculations show that Ce327 has a larger V$_O$ formation energy with lower V$_O$ concentration. Overall, our findings provide valuable insights into the effects of V$_O$ in $R$327 compounds and highlight the potential of Ce327 as a high $T_c$ superconducting material.

**Methods.**

All the lattice relaxation and electronic property calculations for $R$327 are performed using the *ab-initio* VASP package [48-50]. The Perdew-Burke-Ernzerhof (PBE) exchange-correlation functional [51] and projector augmented wave (PAW) pseudopotentials [52] are employed, with the 4$f$ electrons of $R$ atoms treated as core electrons. The lattice and atomic relaxations are made under all considered pressures. Phonon spectra for the Amam and Fmmm phases under variable pressures are calculated using the density functional perturbation theory approach implemented in the PHONONPY software [53-56]. The Wannier downfolding via the Wannier90 code [57] is utilized to obtain on-site energy differences, with La $d$, Ni $d$ and O $p$ orbitals considered. The hopping parameters in the two-orbital model are obtained by downfolding the band structure with Ni-$d_{z2}$ and Ni-$d_{x2-y2}$ orbitals. The on-site Hubbard $U$ is considered with $U$ = 4 eV [58], consistent with the literature [34,59], which is added to the 3$d$ orbitals of Ni.

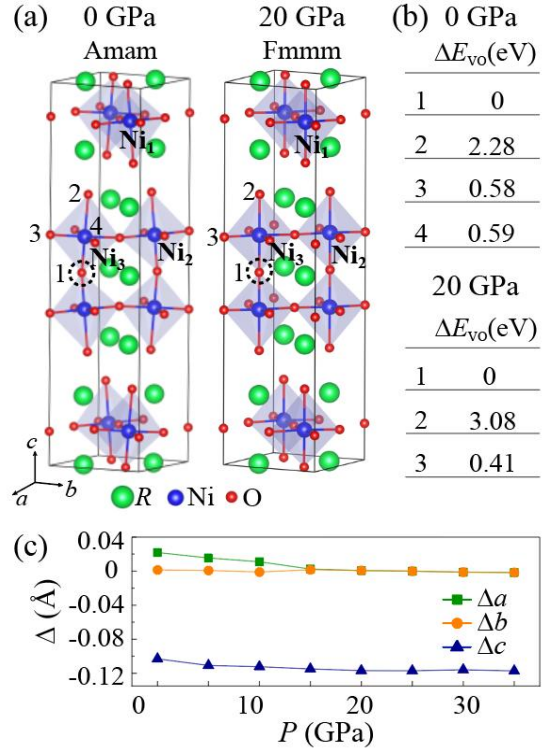

Fig. 1 (a) Schematic crystal structures of $R$327 in the Amam phase (0 GPa, left panel) and Fmmm phase (20 GPa, right panel). The numbers label the nonequivalent positions of oxygen atoms (O$_1$, O$_2$, O$_3$, and O$_4$). Ni$_1$ (Ni atoms without V$_{O1}$), Ni$_2$ (Ni atoms next to the V$_{O1}$) and Ni$_3$ (Ni atoms with V$_{O1}$) represent three nonequivalent Ni atoms when a V$_{O1}$ defect is formed (marked as the dished circles at site 1). (b) Total energy differences for V$_O$ at different positions. Here, the lowest energy one (V$_{O1}$) under each pressure is set as 0. (c) Change of lattice constants under pressure for La327 with the V$_{O1}$ defect.

## Results and Discussions.

**Structures with $V_O$.**

To simplify the analysis and facilitate better comparison with pristine La327, the conventional cells of La327 with total 28 oxygen atoms are selected, leaving the cases for 2×2×1 supercell in Appendix. Similar to previous works, the space group of pristine La327 changes from Amam to Fmmm at pressure around 10 GPa, and the nonequivalent O sites for creating $V_O$ in these two phases are labeled in Fig. 1(a). After full structure relaxations, the total energy difference of La327 with one vacancy is given in Fig. 1(b), in which the most energetically favorable structure under each pressure is set as 0. Interestingly, for both phases, the structures with the loss of intra-bilayer apical octahedra oxygen atom $V_{O1}$ [site 1 in Fig. 1(a)] have the lowest energy. The energies of structures with inter-bilayer apical $V_{O2}$ [site 2 in Fig. 1(a)] are several electron volts higher, while the energies of structures with in-plane $V_{O3}$ and $V_{O4}$ [sites 3 and 4 in Fig. 1(a)] are ~0.5 eV higher. Therefore, it is suggested that intra-bilayer apical $V_{O1}$ most likely exists in experiments. Taking La327 under 20 Gpa as an example, the introduction of intra-bilayer apical $V_{O1}$ gives rise to three non-equivalent Ni atoms, in terms of their distance to the $V_{O1}$: $Ni_1$ is the Ni atom in pristine bilayer $NiO_6$, $Ni_2$ is the Ni atom with corner-shared O connecting to $NiO_5$, and $Ni_3$ is the Ni atom in $NiO_5$ environment. The electronic differences of these Ni atoms will be discussed later. As shown in Fig. 1(c), the loss of apical oxygen atom decreases the lattice constant along $c$-axis largely, while slightly increases the lattice constants in $ab$ plane at 0 GPa. When the pressure is increased, the lattice constants difference between pristine La327 and $V_{O1}$ case changes slightly.

**Electronic properties with apical $V_O$.**

Taking the case of La327 with $V_{O1}$ at 20 GPa as an example, the electronic structures contributed by different Ni atoms are calculated for comparison. Figure 2(a) provides a sketch of the orbital configurations for $Ni_1$ ($Ni_2$) and $Ni_3$ atoms [see Fig. 1(a) for the definition of different Ni atoms]. For the $Ni_1$ atom, the filling of 3$d$ electrons is 7.5 in terms of the formal electron count, giving rise to $Ni^{2.5+}$. The strong hybridization with intra-bilayer apical O splits the $Ni_1$-$d_{z^2}$ to bonding-antibonding molecular-orbital (MO) state. For $Ni_3$ atom around $V_{O1}$, the total 3$d$ filling of is 8.5, resulting in $Ni^{1.5+}$. The distance between two Ni atoms in one bilayer, denoted as $d_{Ni}$, is measured to be 3.38 Å for $Ni_3$-$Ni_3$ atoms, significantly smaller than those of $Ni_1$-$Ni_1$ (~3.82 Å) and $Ni_2$-$Ni_2$ (~3.71 Å). For all Ni atoms, the filling of the in-plane orbitals remains similar, that is, the $t_{2g}$ orbitals are fully occupied, while the $d_{x^2-y^2}$ orbital is still nearly quarter-filled as the case for Ni in La327 without $V_O$.

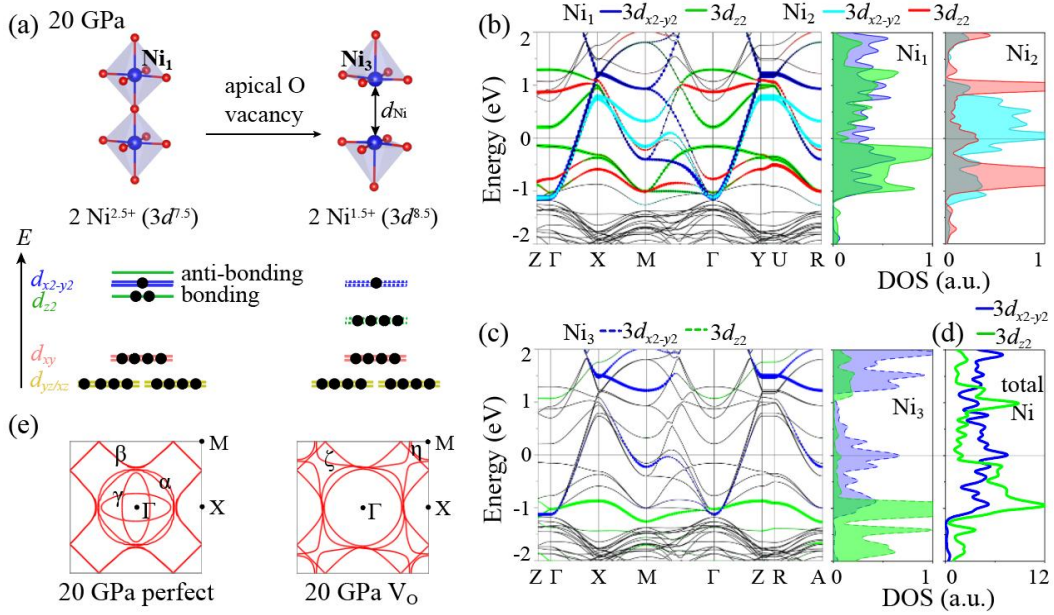

Fig. 2 Electronic structures for La327 with $V_{O1}$ under 20 GPa. (a) Schematic local structures and crystal field splitting for bilayer without (left panel) and with $V_{O1}$ (right panel). The filling of each orbital is indicated by black dots. Here, the two atoms in one bilayer with a total of 15 and 17 electrons for $Ni_1$ and $Ni_3$ are considered. The distance between two Ni atoms in one bilayer is labeled as $d_{Ni}$. (b) Projected band structures and DOS for $Ni_1$ and $Ni_2$ atoms. The $E_F$ is set to zero. (c) Same as (b) but for $Ni_3$ atoms. (d) Projected DOS of Ni $d_{z^2}$ and $d_{x^2-y^2}$ orbitals for all eight Ni atoms. (e) FSs at 20 GPa for La327 without (left panel) and with (right panel) $V_{O1}$. The hole pocket and electron pocket η, as well as FS sheets of α, β and ζ are labeled.

To understand the role of different Ni atoms contributing to the electronic structure of La237, as shown in Figs. 2(b-c), the $d_{x^2-y^2}$ and $d_{z^2}$ orbitals for $Ni_1$, $Ni_2$, and $Ni_3$ are projected. We note that the Ni $t_{2g}$ orbitals are fully filled below the Fermi level. The following differences between three Ni atoms can be observed: 1) For $Ni_1$ atoms, the band dispersions of $d_{z^2}$ and $d_{x^2-y^2}$ orbitals resemble those of Ni atoms in the pristine case (see Fig. 7 in Appendix). However, when the $V_{O1}$ is induced, the inversion symmetry is broken and the energy levels of $Ni_1$ $d_{z^2}$ orbitals are separated from those of $Ni_2$ and $Ni_3$. This leads to a diminished electronic interaction of $Ni_1$ $d_{z^2}$ orbitals between the bilayers and an increased energy splitting of the bonding and antibonding states of $Ni_1$ $d_{z^2}$, as evidenced at the Z and Γ points in Fig. 2(b). 2) For $Ni_2$ atoms, the smaller distance between two $Ni_2$ atoms ($d_{Ni}$) leads to enhanced σ-bond coupling of $Ni_2$ $d_{z^2}$ through O $p_z$ than that of $Ni_1$ $d_{z^2}$. This stronger coupling gives rise to more localized $Ni_2$ $d_{z^2}$ orbital and larger energy splitting of the bonding-antibonding states, which is clearly demonstrated in the projected DOS in the right panel of Fig. 2(b). Moreover, the band dispersion of $Ni_2$ $d_{x^2-y^2}$ reduces largely, leading to a large proportion around $E_F$. 3) For $Ni_3$ atoms, with the large direct σ-bond coupling, the $Ni_3$ $d_{z^2}$ orbitals are fully filled and experience a significant energy downwards shifting, approaching the energy levels of the $t_{2g}$ orbitals. Meanwhile, the energy of $Ni_3$ $d_{x^2-y^2}$ shift upwards slightly, accompanied by a small decreasing in orbital filling. Therefore, the inclusion of $V_{O1}$ significantly renormalize the electronic structure around the Fermi level.

Figure 2(d) also gives the sum of DOS for all eight Ni atoms in La327 with $V_{O1}$. Compared to the large DOS of Ni $d_{z^2}$ bonding states across $E_F$ for the pristine case [see Fig. 7 (a) in Appendix], the Ni $d_{z2}$ shifts away and has much smaller proportion of around $E_F$ for $V_O$ case. Meanwhile, the proportion of $d_{x2-y2}$ orbital around $E_F$ increases significantly, even surpassing the proportion of $d_{z2}$ state. This contrasts with the pristine case, where $d_{z2}$ has a much larger orbital proportion compared to $d_{x2-y2}$. As a result, the total DOS of Ni 3d orbital around $E_F$ decreases with the inclusion of $V_O$. Given that the bonding state of Ni $d_{z2}$ around Fermi level may be important for the appearance of superconductivity in La327 under pressure, the inclusion of $V_{O1}$ may be detrimental to the superconductivity. In particular, the inhomogeneous distribution of $V_{O1}$ may induce the inhomogeneous superconducting properties in the samples. In addition, the Fermi surfaces (FSs) of La327 without and with $V_{O1}$ are illustrated in Fig. 2(e) for comparison. For the pristine case, the observed hole pocket γ (primarily composed by Ni $d_{z2}$ orbital) as well as the FS sheets α and β (a mixture of $d_{z2}$ and $d_{x2-y2}$ orbitals) are consistent with the results in previous works [15,59]. When $V_{O1}$ exist, the FS sheets α and β remain largely unchanged while the hole pocket γ diminishes because of the strong downwards shift of $d_{z2}$. Additionally, the electron pocket η around M point and FS sheet ζ emerge, which are mainly composed of $d_{x2-y2}$ states originated from $Ni_2$ and $Ni_3$ atoms.

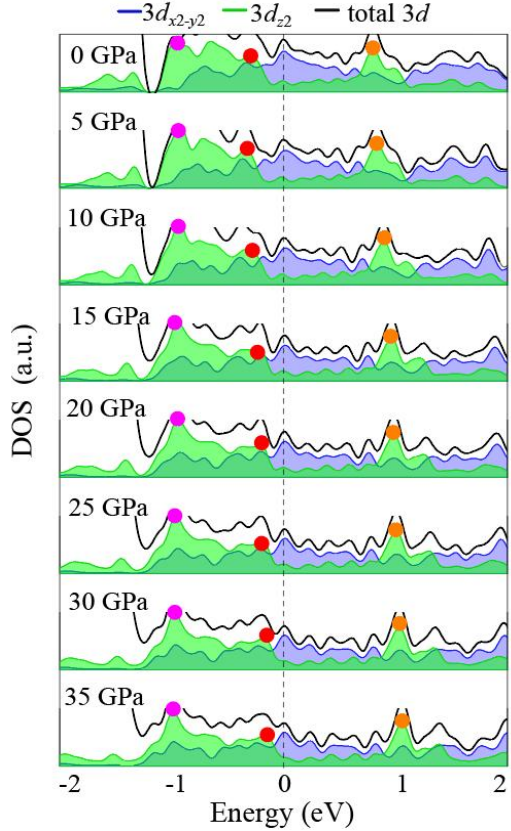

Fig. 3 Projected DOS of $d_{z2}$ and $d_{x2-y2}$ orbitals for the sum of eight Ni atoms in La327 with $V_{O1}$ as a function of external pressure. Red, magenta and orange points represent the peak positions contributed mainly by the bonding $Ni_1$ $d_{z2}$ states, bonding $Ni_2$ $d_{z2}$/$Ni_3$ $d_{z2}$ states, and antibonding $Ni_2$ $d_{z2}$ states, respectively.

It is interesting to further understand the evolution of electronic properties of La327 with $V_O$ under increased pressures, which is presented in Fig. 3. The projected DOS represents the combined contribution of all eight Ni atoms in the unit cell. As the pressure increases, the bonding state of Ni$_1$-$d_{z^2}$ shifts upward (indicated by red points), similar to the behavior observed in pristine La327. In contrast, the energy level of bonding state of Ni$_2$ $d_{z^2}$ remains almost unchanged (shown by magenta points), while the energy level of the anti-bonding state of Ni$_2$ $d_{z^2}$ increases with increasing pressure (represented by orange points). The shape of $d_{x^2-y^2}$ orbital undergoes minimal changes around the $E_F$ as the pressure increases, while $E_F$ always cross the peak position of $d_{x^2-y^2}$ orbital for all pressures.

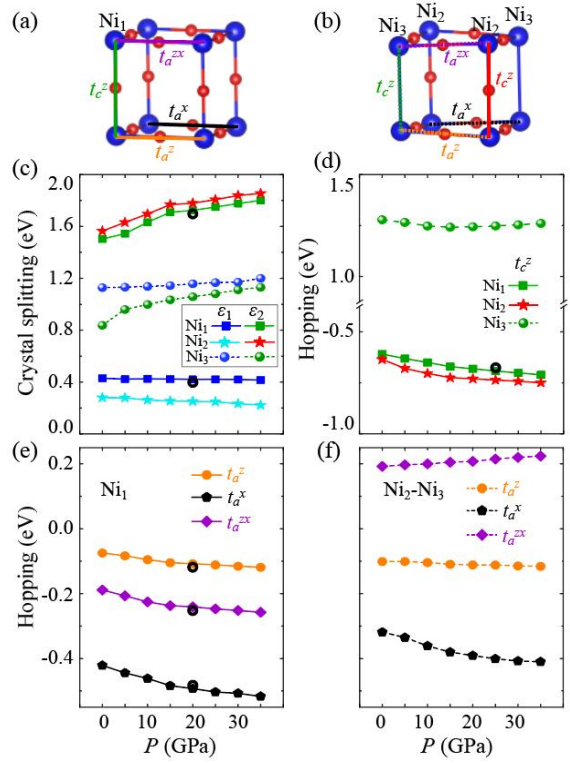

Fig. 4 On-site energy differences and hopping parameters for three non-equivalent Ni atoms in La327 with V$_{O1}$. Sketch of the Ni bilayer structures (a) without and (b) with V$_{O1}$. The hopping parameters are labeled by colored lines: $t_c^z$ (green, coupling between NN intra-bilayer $d_{z^2}$ along $c$-axis), $t_a^z$ (orange, coupling between NN $d_{z^2}$ along $a$-axis), $t_a^x$ (black, coupling between NN $d_{x^2-y^2}$ along $a$-axis) and $t_a^{zx}$ (purple, coupling between NN $d_{z^2}$ and $d_{x^2-y^2}$ along $a$-axis). (c) Crystal-field splitting $\varepsilon_1$ (between $d_{x^2-y^2}$ and $d_{z^2}$ orbitals) and $\varepsilon_2$ (between $d_{z^2}$ and $t_{2g}$ orbitals) for three nonequivalent Ni atoms as a function of pressure. (d) $t_c^z$ as a function of pressure. (e) $t_a^z$, $t_a^x$, and $t_a^{xz}$ for Ni$_1$ atoms as a function of pressure. (f) Same as (e) but for the hoppings between the orbitals of Ni$_2$ and Ni$_3$ atoms. The black circles in (c-e) denote the corresponding values for Ni atoms in pristine La327 under 20 GPa, which are close to the corresponding symbols of Ni$_1$.

**Orbital interactions with V$_O$.**

The minimal two-orbital model, consisting of $d_{z^2}$ and $d_{x^2-y^2}$ orbitals, has been proposed to describe the electronic properties of pristine La327. It has been found that, in addition to the coupling between $d_{z^2}$ orbitals along $c$-axis ($t_c^z$), significant hybridization between $d_{z^2}$ and itinerant $d_{x^2-y^2}$ orbitals along $a$-axis ($t_a^{zx}$) may play a crucial role in achieving high-$T_c$ superconductivity [13,15,21,22]. To investigate the effects of V$_O$ on these orbital interactions, Wannier downfolding analysis is performed to examine the changes in the model parameters, and the Wannier orbitals are given in Fig. 8 in Appendix. Figures 4(a) and (b) provide a schematic representation of the Ni bilayer structures, with a noticeable lattice distortion observed in Fig. 4(b). As shown in Fig. 4(c), the crystal-field splitting for Ni$_1$ and N$_2$ atoms are fairly close but both are very different from that of Ni$_3$. Regarding the energy difference between $d_{x2-y2}$ and $d_{z2}$ orbitals ($\varepsilon_1$), it is observed that Ni$_2$ has a lower value compared to Ni$_1$ because of the downward energy shift of Ni$_2$ $d_{x2-y2}$ orbital [Fig. 2(b)]. As pressure increases, $\varepsilon_1$ for both Ni$_1$ and Ni$_2$ decreases, attributing to the upward shift of the $d_{z2}$ orbital. On the other hand, for Ni$_3$, which has fully occupied $d_{z2}$ states, $\varepsilon_1$ is significantly larger than that in Ni$_1$ and Ni$_2$. With increasing pressure, $\varepsilon_1$ for Ni$_3$ slightly increases due to the downward shift of the $d_{z2}$ orbital. The energy difference between $d_{z2}$ and $t_{2g}$ orbitals ($\varepsilon_2$) is slightly higher for Ni$_2$ compared to Ni$_1$, and both values increase with pressure. However, the $\varepsilon_2$ values for Ni$_3$ are much smaller, even smaller than the $\varepsilon_1$ values for Ni$_3$. In addition, the loss of oxygen atom has minimal impact on the orbital alignment of Ni$_1$ compared to the Ni atom in pristine La327, which can be inferred from the similar projected band structures [Fig. 2(b) and Fig. 7] as well as the comparable model parameters.

Next, the primary orbital interactions in the two-orbital model are discussed. The out-of-plane nearest-neighbor (NN) $\sigma$-bonding coupling $t_c^z$ is presented in Fig. 4(d). Due to the reduced distance of the bilayer $d_{z2}$ orbitals, $t_c^z$ for Ni$_2$ is slightly larger than that of Ni$_1$, and both values increase with increasing pressure. Surprisingly, without the apical oxygen atom, the orbital coupling between Ni$_3$-$d_{z2}$ orbitals significantly increases with a positive $t_c^z$ value between two positively ionized Ni$_3$. And the $d_{z2}$-like Wannier orbitals are depicted in Figure 8 (e), revealing that the lose of oxygen atoms results in the presence of residual electrons remaining at the apical oxygen site.

Regarding the in-plane orbital coupling shown in Fig. 4(e), Ni$_1$ exhibits similar values to those of the Ni atom in pristine La327 (black circles). Apart from the substantial intra-orbital coupling of $d_{x2-y2}$ ($t_a^x$), the inter-orbital coupling between $d_{x2-y2}$ and $d_{z2}$ ($t_a^{zx}$) is considerable, while the intra-orbital coupling of $d_{z2}$ ($t_a^z$) is relatively small. All these coupling strengths increase under a higher pressure. In the presence of V$_{O1}$, Ni$_3$ is adjacent to Ni$_2$ [Fig. 4(b)], resulting in in-plane orbital coupling between them, whose values are given in Fig. 4(f). The coupling between $d_{z2}$ of Ni$_2$ and Ni$_3$ along the $a$-axis ($t_a^z$) is similar to the value observed for Ni$_1$-$d_{z2}$, while the coupling between Ni$_2$-$d_{x2-y2}$ and Ni$_3$-$d_{x2-y2}$ ($t_a^x$) is relatively smaller compared to Ni$_1$-$d_{x2-y2}$. The most dramatic difference is the change in the intra-orbital coupling between Ni$_2$-$d_{z2}$ and Ni$_3$-$d_{x2-y2}$ (also between Ni$_2$-$d_{x2-y2}$ and Ni$_3$-$d_{z2}$) along $a$-axis ($t_a^{zx}$). This coupling strength has nearly absolute value to that of Ni$_1$ case but with a completely opposite sign. Moreover, as the case in Ni$_1$, in-plane couplings between orbitals of Ni$_2$ and Ni$_3$ atoms also increase with pressure. We note that a recent theoretical study has assumed zero intra-bilayer coupling between Ni$_3$-$d_{z2}$ ($t_c^z$) with the loss of apical oxygen. This assumption could lead to improved Fermi surface nesting and consequently reduced interaction strength [24]. However, our results demonstrate an increased, but opposite $t_c^z$ value in the V$_O$ case [Fig. 4(d)]. Therefore, further investigation is necessary to explore how the changes in orbital interactions within this two-orbital model affect superconducting properties.

**Formation energies of V$_O$.**

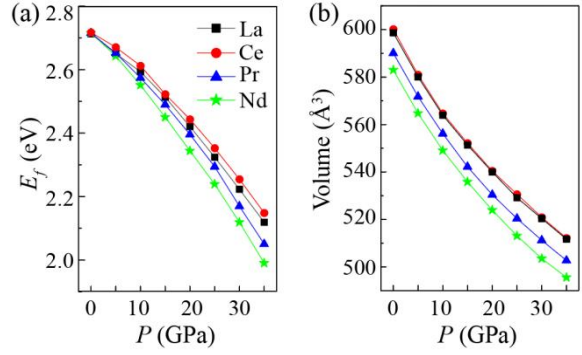

Fig. 5 (a) Formation energies $E_f$ of apical $V_{O1}$ for $R$327 ($R$ = La, Ce, Pr, Nd) under pressure. (b) The change of volume for pristine $R$327.

In the above discussions, we explore the potential position and impact of $V_{O1}$ on electronic properties under the assumption that $V_{O1}$ exists in La327, which could be true in the experiments [6-9]. Meanwhile, it is expected that the possibility to form $V_{O1}$ can vary significantly as lattice constants change under different pressures. The formation energies ($E_f$) of single $V_{O1}$ in the La327 conventional cell can be evaluated as:

$$E_f = E[\text{La}_{12}\text{Ni}_8\text{O}_{27}] + \mu[\text{O}] - E[\text{La}_{12}\text{Ni}_8\text{O}_{28}] \tag{1},$$

where $E[\text{La}_{12}\text{Ni}_8\text{O}_{27}]$ and $E[\text{La}_{12}\text{Ni}_8\text{O}_{28}]$ are the total energies of $\text{La}_{12}\text{Ni}_8\text{O}_{27}$ and $\text{La}_{12}\text{Ni}_8\text{O}_{28}$, respectively. $\mu[\text{O}]$ is set to the chemical potential of 1/2 $O_2$. In general, the lower the $E_f$, the more likely to form $V_{O1}$.

Figure 5(a) illustrates the change of $E_f$ for $V_{O1}$ under pressure. It is evident that $E_f$ significantly decreases with pressure, indicating that the formation of $V_{O1}$ becomes more likely under high pressure. This can be mainly attributed to the significant reduction in lattice volume as shown in Fig. 5(b), favoring the formation of vacancy defects. As discussed above, due to the detrimental effect of $V_{O1}$ to superconductivity, the structures with fewer $V_O$ may be preferred. Considering the larger volume and $c/a$ value, Ce327 is expected to have a higher $E_f$ of $V_{O1}$ compared to La327. Indeed, as depicted in Fig. 5(a), $E_f$ values for Ce327 are always higher than those for La327, and this difference becomes more pronounced at higher pressures. In addition, the comparison calculations for Pr327 and Nd327 are also conducted, revealing smaller volume and thus smaller $E_f$ than La327. The comparable electronic properties and FS (as shown in Fig. 7) for pristine La327 and Ce327, along with the similar pairing strength calculated by the multi-orbital random phase approximation (RPA) model [34], indicate the potential for superconductivity in Ce327. That's to say, Ce327 could be a promising candidate for achieving nickelate-based high $T_c$ superconductors with potentially lower $V_O$ concentration.

**Electronic properties of Ce327 with $V_O$.**
Similar to La327, Ce327 undergoes a phase transition from Amam to Fmmm at ~11 GPa [34]. Figure 6(a) depicts the phonon spectrum of Ce327 in the Fmmm phase at 20 GPa. The absence of imaginary frequencies suggests its thermodynamic stability. The band structure is shown in Fig. 6(b) (left panel), with the $d_{x2-y2}$ and $d_{z2}$ orbitals projected for all these eight Ni atoms. Figure 6(b) right panel displays the projected DOS for three nonequivalent Ni atoms. The definition of $Ni_1$, $Ni_2$, and $Ni_3$ are same as those in La327 [Fig. 1(a)]. Interestingly, it is evident that the band structure of Ce327 closely resembles that of La327. The presence of $V_{O1}$ has a slight influence on the orbital dispersion of $Ni_1$ but causes a downshift in the $d_{z^2}$ orbital away from the $E_F$. Additionally, the $d_{z^2}$ orbital for $Ni_2$ becomes more localized, while the DOS of $d_{x^2-y^2}$ around $E_F$ exhibits a significant increase. For the $Ni_3$ atom, the $d_{z^2}$ orbital is fully occupied, accompanied by a slight upward shift of the $d_{x^2-y^2}$ orbital. The vanishing of γ pocket along with the appearance of hole pocket η and FS sheet ζ [Fig. 6(c)] are same to the case in La327 with $V_{O1}$ [Fig. 2(e)]. The electronic properties of Pr327 and Nd327 with $V_{O1}$ are presented in Fig. 9 in Appendix, which exhibit similar features to those observed in La327.

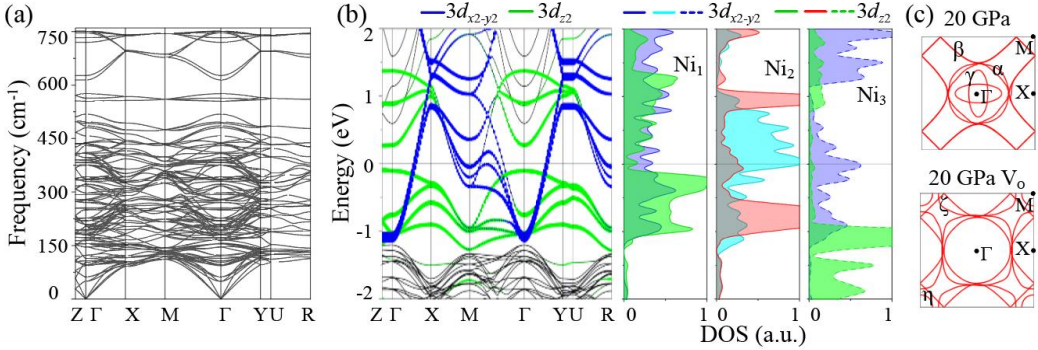

Fig. 6 (a) Phonon spectrum for pristine Ce327 under 20 GPa. (b) Projected band structure with $d_{x^2-y^2}$ and $d_{z^2}$ for all Ni atoms (left panel) and projected DOS for three non-equivalent Ni atoms separately (right panel) for Ce327 under 20 GPa with $VO_1$. (c) FSs for Ce327 without (upper panel) and with $V_{O1}$ (left panel).

**Discussion and Conclusion.**

It is worth noting that the concentration of $V_O$ considered in our study is relatively high ($R_3Ni_2O_{6.75}$, $\delta = 0.25$). Meanwhile, we also consider that the situations at a lower $V_O$ concentration, as demonstrated by the DOS for 2×2×1 supercell in Fig. 10 in Appendix (corresponding to $R_3Ni_2O_{6.9375}$, $\delta = 0.0625$). The main features for each Ni atom are comparable to those observed in $R_3Ni_2O_{6.75}$. Specifically, the sum of DOS for $d_{z^2}$ around $E_F$ decreases, while the DOS for $d_{x^2-y^2}$ slightly increases, as the case in $R_3Ni_2O_{6.75}$.

In conclusion, our first-principles calculations demonstrate that the most energetically favorable site for the formation of oxygen vacancy in $R$327 is the apical octahedral oxygen connecting $NiO_6$ bilayers. The formation of oxygen vacancy results in the increased filling of Ni $d_{z^2}$ orbital, which greatly reduces its proportion at $E_F$. Interestingly the orbital hopping between nearest-neighbour $d_{z^2}$ orbitals along $c$-axis increases significantly with opposite sign for $V_O$ case. Furthermore, the inter-orbital coupling between $d_{z^2}$ and $d_{x^2-y^2}$ also changes sign but with a comparable strength as the case without $V_O$. Therefore, the changes of Ni $d_{z^2}$ orbitals via $V_O$ may be harmful to the superconductivity in the $R$327, given that the Ni $d_{z^2}$ orbitals around the Fermi level are critical to its superconductivity. Finally, holding a larger lattice volume but similar electronic properties, we suggest that Ce327 is a promising candidate for high $T_c$ superconductivity with lower $V_O$ concentrations.

**Acknowledgments**

This work is supported by the National Key Research and Development of China (Grant No. 2022YFA1402401), NSFC (Grant No. 12088101), and NSAF (Grant No. U2230402). Computations are done at the Tianhe-JK supercomputer at CSRC.